\documentclass[lettersize,journal]{IEEEtran}
\usepackage{amsmath,amsfonts}
\usepackage{amsthm}
\usepackage{amssymb}
\theoremstyle{definition}

\newtheorem{theorem}{Theorem}

\usepackage[inline]{enumitem}
\usepackage{makecell} 
\usepackage{bbm}
\usepackage[ruled,linesnumbered]{algorithm2e}
\usepackage{array}
\usepackage{subfigure}
\usepackage{textcomp}
\usepackage{stfloats}
\usepackage{diagbox}
\usepackage{booktabs}
\usepackage{multirow}
\usepackage{makecell} 
\usepackage{url}
\usepackage{verbatim}
\usepackage{graphicx}
\usepackage{cite}
\usepackage[colorlinks,
            linkcolor=blue,
            anchorcolor=blue,
            citecolor=blue,
            urlcolor=black]{hyperref}
\usepackage{orcidlink}

\begin{document}

\title{Brownian Bridge Diffusion-Assisted Joint Channel Estimation and Data Detection for Jamming-Resilient OFDM Receivers}

\author{Honghan She,~\IEEEmembership{Graduate Student Member, IEEE}, Yufan Cheng, Siya Huang, \\Pengyu Wang,~\IEEEmembership{Member, IEEE}, Tieming Sun, Kaikai Yang
\thanks{This work was supported by the National Natural Science Foundation of China under Grants 62131005 and 62071096, by the Fund Project of National Key Laboratory of Wireless Communications under grant 2024-kgr-JJ-04, by the Fundamental Research Funds for the Central Universities under Grant 2242022k60006, and by the Enterprise Cooperation Project under Grant 250989. (\textit{Corresponding author: Yufan Cheng}).}
\thanks{Honghan She, Yufan Cheng, Siya Huang, Tieming Sun, and Kaikai Yang are with the National Key Laboratory of Wireless
Communications, University of Electronic Science and Technology of
China, Chengdu 611731, China (e-mail: \href{mailto:hhshe2000@gmail.com}{hhshe2000@gmail.com}, \href{mailto:chengyf@uestc.edu.cn}{chengyf@uestc.edu.cn}).}
\thanks{Pengyu Wang is with School of Artificial Intelligence, University of Science and Technology Beijing, Beijing 100083, China (e-mail: \href{mailto:wangpengyu@ustb.edu.cn}{wangpengyu@ustb.edu.cn}).}
}


\maketitle

\begin{abstract}
In next-generation wireless networks, the growing density of devices and limited spectrum resources pose severe jamming challenges to fragile legitimate communication links in the wireless electromagnetic environment. Crucially, when jamming overlaps with pilot and data symbols in both time and frequency domains, it inflicts a severe bottleneck on receiver-side joint estimation and detection. Existing schemes often lack an effective framework to combat such jamming contamination, thereby failing to guarantee reliable transmission. To address this issue, we propose a Brownian bridge diffusion-assisted joint channel estimation and data detection (BBD-JCED) framework for jamming-resilient orthogonal frequency-division multiplexing (OFDM) receivers. The proposed framework consists of a jamming suppression module and a Brownian bridge-joint recovery module. The former identifies jamming-contaminated bins in the short-time Fourier transform (STFT) domain and suppresses the dominant jamming components through spectral notching. The latter establishes a Brownian bridge-based probabilistic graph to model the probabilistic transition process from the suppressed OFDM observation to the coded bits, and performs a reverse recovery process to enable enhanced joint channel estimation and data detection. Moreover, we derive a variational lower bound (VLB) to reformulate the intractable bit estimate problem into a tractable training objective, providing a probabilistic foundation for the loss function design. Furthermore, we develop a two-stage training algorithm to separately optimize the jamming suppression and Brownian bridge-joint recovery modules. Simulation results under representative jamming types demonstrate that the proposed framework achieves superior jamming suppression and bit recovery performance over the considered baseline schemes while maintaining competitive inference complexity.
\end{abstract}

\section{Introduction}
\IEEEPARstart{W}{ith} the evolution of beyond fifth-generation (B5G) and sixth-generation (6G) mobile networks, increasingly dense and heterogeneous wireless deployments are intensifying spectrum coexistence and interference \cite{ref:P.L, ref:T.O}. The resulting electromagnetic environment not only degrades quality of service but also increases the vulnerability of legitimate links to malicious jamming \cite{ref:L.C}. By directly reducing the signal-to-jamming-plus-noise ratio (SJNR) at the receiver, jamming attacks can corrupt physical-layer observations in a manner that upper-layer security mechanisms cannot readily compensate for \cite{ref:H.P, ref:K.N.V, ref:P.W-1}. Therefore, the design of jamming-resilient physical-layer receivers has become an important requirement for reliable B5G/6G communications.

Traditional anti-jamming techniques, such as frequency hopping spread spectrum (FHSS) and direct sequence spread spectrum (DSSS), provide resistance by spreading or varying the occupied spectrum, but fixed transmission patterns are inadequate for adapting to reactive jammers in dynamic spectrum environments \cite{ref:X.W-1}. In conventional orthogonal frequency-division multiplexing (OFDM) receivers, least squares (LS) or minimum mean-square error (MMSE) algorithms are typically used for pilot-based channel state information (CSI) estimation \cite{ref:H.L.V.T}, followed by linear or nonlinear equalization and data detection \cite{ref:J.G.P}. When jamming overlaps with both pilot and data resources in the time-frequency domain, the resulting CSI estimation bias propagates into subsequent equalization and detection, causing severe error propagation. Hence, a jamming-resilient receiver should not only suppress the interfering components but also preserve sufficient information for reliable channel estimation and data detection under residual jamming and imperfect CSI.

Deep learning has provided new tools for improving receiver processing under complex channel conditions \cite{ref:P.W-2, ref:P.W-3}. For channel estimation, architectures such as ChannelNet \cite{ref:M.S}, ReEsNet \cite{ref:L.L}, HA02 \cite{ref:D.L}, AE-DENet \cite{ref:E.F}, and AdaFortiTran \cite{ref:B.G} learn to interpolate pilot observations into the full channel response. To exploit the coupling between channel estimation and data detection, end-to-end receivers have also been developed, including the deep neural network (DNN) receiver in \cite{ref:H.Y}, the decoder and encoder convolutional neural network (DECNN) in \cite{ref:J.L}, and TransDetector in \cite{ref:Y.L}. In parallel, model-driven designs such as ComNet \cite{ref:X.G} and the channel estimation network (CENet)/channel conditioned recovery network (CCRNet) architecture in \cite{ref:X.Y} combine conventional receiver knowledge with neural networks. Although these discriminative approaches improve channel estimation and detection under their target operating conditions, they primarily learn direct mappings from impaired observations to point estimates or decisions. Under strong or non-stationary jamming, however, pilot and data observations can be heavily contaminated or partially unreliable, making direct discriminative recovery increasingly difficult.

Generative artificial intelligence (GAI) offers an alternative by learning the distribution of legitimate signals and using it as a prior for recovering severely corrupted observations. Generative adversarial networks (GANs) have been used to complete missing spectral information \cite{ref:H.H}, while variational autoencoders (VAEs) have been investigated for data symbol reconstruction under strong jamming \cite{ref:I.W.C.W}. More recently, diffusion models (DMs) have attracted attention because of their strong distribution-modeling capability \cite{ref:D.F, ref:C.Z}. Conventional DMs define a forward perturbation process from target data toward a tractable prior distribution, typically Gaussian noise, and learn the corresponding reverse denoising process. They have been applied to jamming source separation \cite{ref:X.W-2, ref:J.O}, score-based separation of digital communication signals \cite{ref:T.J}, high-dimensional channel estimation \cite{ref:X.Z}, and signal reconstruction following jamming suppression or notching \cite{ref:M.Y, ref:H.S, ref:H.S-TVT}. Diffusion bridge models constitute a related but distinct class of diffusion-based generative processes: instead of connecting data to an independent noise prior \cite{ref:J.H, ref:J.S}, they construct a stochastic process conditioned on prescribed endpoint states or distributions. A Brownian bridge is a specific diffusion bridge obtained by conditioning Brownian motion on prescribed target and terminal endpoints, and Brownian bridge diffusion (BBD) uses this Brownian-bridge reference process to model stochastic evolution between two prescribed endpoints. BBD has been studied for image-to-image translation \cite{ref:B.L, ref:K.Z} and has further been introduced into physical-layer channel estimation and uplink denoising tasks \cite{ref:S.X, ref:Z.H}. Another diffusion-bridge formulation was adopted by the diffusion bridge receiver network (DBRNet) \cite{ref:R.Z} for joint channel estimation and data detection in superimposed-pilot OFDM systems, where the received-signal distribution and the joint distribution of the transmitted data and channel impulse response are specified as the two endpoint distributions.

Despite these advances, a task-specific gap remains in diffusion-based receivers for jamming-resilient joint channel estimation and data detection. Conventional noise-to-data DMs do not explicitly exploit the available jamming-suppressed observation as an endpoint of the generative trajectory. Existing diffusion bridge methods provide endpoint-constrained stochastic evolution, but their endpoint and conditioning designs are generally developed for other reconstruction tasks and do not explicitly account for the coupled effects of residual jamming and imperfect CSI in OFDM reception. This motivates a BBD-assisted receiver that directly constructs a probabilistic recovery path between the jamming-suppressed OFDM observation and the coded bits, while incorporating the refined CSI and estimated jamming information as physical conditional guidance. In this manner, channel refinement, symbol recovery, and soft data detection are coupled within a unified receiver, forming the core methodology of the proposed jamming-resilient joint channel estimation and data detection framework. The main contributions of this paper are summarized as follows:
\begin{itemize}
\item 
We propose a Brownian bridge diffusion-assisted joint channel estimation and data detection (BBD-JCED) framework for jamming-resilient OFDM receivers. The framework consists of a jamming suppression module and a Brownian bridge-joint recovery module. The former identifies contaminated time-frequency bins in the short-time Fourier transform (STFT) domain and suppresses the dominant jamming components through spectral notching. The latter performs a reverse recovery process from the suppressed OFDM observation toward the coded bit, enabling enhanced joint channel estimation and data detection.
\item 
We establish a Brownian bridge-based probabilistic graph to model the probabilistic transition process from the suppressed OFDM observation to the coded bits for the Brownian bridge-joint recovery module, while the refined CSI and estimated spectral notching mask serve as physical side information. Based on this, we derive a variational lower bound (VLB) to reformulate the intractable bit estimate problem into a tractable training objective, providing a probabilistic foundation for the loss function design of the proposed framework.
\item 
We develop a two-stage training algorithm tailored to the proposed framework. The jamming suppression module is optimized first to identify and suppress jamming samples in the STFT domain. Then, its parameters are frozen, and the Brownian bridge-joint recovery module is optimized using the VLB-guided objective that jointly accounts for CSI refinement, bridge endpoint recovery, and soft bit detection. Simulation results under multiple representative jamming types demonstrate that the proposed framework achieves improved jamming suppression and bit recovery performance over the considered baseline schemes with competitive inference complexity.
\end{itemize}

The remainder of this paper is organized as follows. Section \ref{model} presents the jamming scenario and signal model. Section \ref{method} details the proposed BBD-JCED framework. Extensive simulation results and performance evaluations are presented in Section \ref{result}. Finally, Section \ref{conclusion} concludes the paper.

\textit{Notations:} Column vectors and matrices are denoted by bold lowercase and uppercase letters, respectively: $\mathbf{a}$ and $\mathbf{A}$. Sets and mappings are denoted by fancy uppercase letters: $\mathcal{A}$. $\|\cdot\|_2$, $\|\cdot\|_F$ and $(\cdot)^{\mathrm{H}}$ denote the 2-norm, Frobenius norm and conjugate transpose, respectively. $\langle\cdot,\cdot\rangle_F$ denotes the Frobenius inner product of two matrices. $\mathbb{C}$ denotes the set of complex numbers. $\mathbb{R}_{>0}$ denotes the set of positive real numbers. $\mathbb{N}$ denotes the set of positive natural numbers. $\odot$ denotes the element-wise product. $\Re$ and $\Im$ denote the real and imaginary parts, respectively. $\nabla$ and $\int$ denote the gradient operator and the integration operation, respectively. $\mathbb{E}[\cdot]$ denotes the expectation operator. $\mathbb{I}(\cdot)$ denotes the indicator function. $D_{\text{KL}}(\cdot,\cdot)$ denotes the Kullback-Leibler (KL) divergence between two variables. $\mathcal{N}(\mathbf{0}, \mathbf{I})$ denotes the zero-mean Gaussian distribution with identity covariance $\mathbf{I}$, indicating independent and identically distributed noise across all dimensions.

\section{System Model}
\label{model}

We first describe the wireless jamming scenario, followed by a mathematical formulation of the corresponding signal model.

\subsection{Wireless Jamming Scenario}

\begin{figure}[!h]
    \centering
    \includegraphics[width=0.32\textwidth]{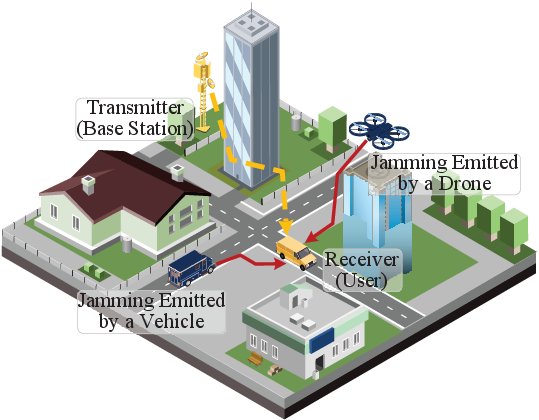}
    \caption{Wireless jamming scenario. The yellow dashed line represents the communication link from the transmitter to the receiver, while the red solid line represents the jamming link from a vehicle or a drone to the receiver.}
    \label{fig:scenario}
\end{figure}

Fig. \ref{fig:scenario} illustrates the wireless jamming scenario. The yellow dashed line denotes the communication link from the transmitter to the receiver, which facilitates data transmission between the two legitimate parties. The red solid line represents the jamming link, originating from a ground vehicle or an airborne drone, targeting the legitimate receiver and degrading the quality of the communication link.

\subsection{Signal Model}

\begin{figure}[!ht]
    \centering
    \includegraphics[width=0.42\textwidth]{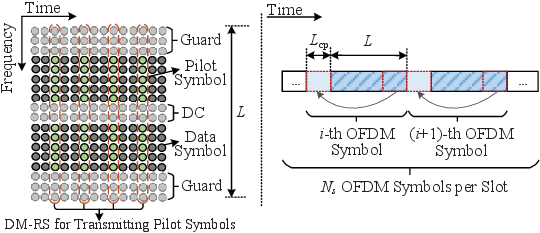}
    \caption{Exemplary OFDM time-frequency grid and time-domain waveform.}
    \label{fig:ofdm}
\end{figure}

Consider an orthogonal frequency-division multiplexing (OFDM) communication link, with an exemplary OFDM structure depicted in Fig. \ref{fig:ofdm}. An OFDM symbol comprises $L$ subcarriers, where $L$ is the size of the fast Fourier transform (FFT)/inverse FFT (IFFT). According to the physical downlink shared channel (PDSCH) in 3GPP TS 38.211 \cite{ref:3gpp}, among these $L$ subcarriers, a total of $L_{\text{null}}$ nullified guard subcarriers are placed at the center (i.e., DC guard band) and at the edges (i.e., edge guard bands), while the remaining useful subcarriers are allocated for data or pilot symbols. For OFDM symbols configured as demodulation reference signals (DM-RS), these useful subcarriers exclusively carry pilot symbols. On the one hand, all data symbols are generated by $Q$-ary constellation modulation (e.g., quadrature phase shift keying (QPSK)) of encoded bits $\mathbf{b}\in\{0,1\}^{N_b}$, where $\mathbf{b}$ is produced by channel coding (e.g., low-density parity-check (LDPC) code) of the data bits. On the other hand, the pilot symbols are a priori known to both the transmitter and receiver. The frequency-domain symbols on the $L$ subcarriers are transformed into time-domain samples via the IFFT, and a cyclic prefix (CP) is appended to form an OFDM symbol of length $L_{\text{CP}}+L$, where $L_{\text{CP}}$ denotes the CP length. Furthermore, $N_s$ OFDM symbols constitute an OFDM slot waveform, which serves as the transmitted signal $\mathbf{s}\in\mathbb{C}^{N}$ in this work, where $N=N_s(L_{\text{CP}}+L)$.

\begin{figure}[!ht]
    \centering
    \subfigure[CSN jamming with $4$ combs]{
        \centering
        \includegraphics[width=0.195\textwidth]{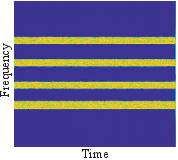}
    }
    \subfigure[LFM jamming with $4$ periods]{
        \centering  
        \includegraphics[width=0.195\textwidth]{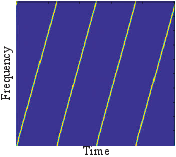}
    }
    \caption{Time-frequency spectrum of CSN jamming and LFM jamming.}
\label{fig:tf_jam}
\end{figure}

The received signal in a jammed slot can be mathematically expressed as
\begin{equation}
    \begin{aligned}
    \mathbf{y}=\mathbf{H}_t\mathbf{s} + \mathbf{G}_t\mathbf{w}+\mathbf{n},
    \end{aligned}
\label{eq:rx_signal}
\end{equation}
where $\mathbf{y}\in\mathbb{C}^{N}$ is the received OFDM time-domain signal, $\mathbf{H}_t\in \mathbb{C}^{N\times N}$ and $\mathbf{G}_t\in \mathbb{C}^{N\times N}$ denote the matrix representations of the channel impulse responses (CIR) for the communication and jamming links, respectively, $\mathbf{n}\in \mathbb{C}^{N}$ is the additive white Gaussian noise (AWGN), and $\mathbf{w}\in \mathbb{C}^{N}$ represents the time-domain jamming samples. Typical jamming types include comb-spectrum noise (CSN) jamming and linear frequency modulation (LFM) jamming. The former can be viewed as the superposition of partial band noise (PBN) from jammers in the frequency domain, while the latter corresponds to a sweep signal from jammers that varies linearly with time \cite{ref:P.W-4}. Fig. \ref{fig:tf_jam} illustrates the time-frequency spectrum of CSN and LFM jamming.

Jamming suppression in the STFT domain has been shown to effectively mitigate the impact of jamming samples on the received signal \cite{ref:H.S}, making it a critical component of jamming-resilient OFDM receivers. Let $\mathcal{F}: \mathbb{C}^N\rightarrow\mathbb{C}^{K\times R}$ and $\mathcal{F}^{-1}: \mathbb{C}^{K\times R}\rightarrow\mathbb{C}^N$ denote the STFT and its inverse, with $K$ frequency bins and $R$ time bins. Let $\hat{\mathbf{M}}\in\{0,1 \}^{K\times R}$ denote the spectral notching mask, where zero elements indicate the positions of jamming samples. In this work, $\hat{\mathbf{M}}$ is estimated via a jamming detector. Consequently, the suppressed signal $\tilde{\mathbf{y}}\in\mathbb{C}^N$ can be expressed as
\begin{equation}
    \begin{aligned}
        \tilde{\mathbf{y}}&=\mathcal{F}^{-1}(\hat{\mathbf{M}}\odot\mathcal{F}(\mathbf{y})) \\
        &=\mathbf{U}\mathbf{H}_t\mathbf{s}+\mathbf{U}\mathbf{G}_t\mathbf{w}+\mathbf{U}\mathbf{n},
    \end{aligned}
\label{eq:js}
\end{equation}
where $\mathbf{U}\in\mathbb{C}^{N\times N}$ is the matrix representation of the composite mapping $\mathcal{F}^{-1}(\hat{\mathbf{M}}\odot\mathcal{F}(\cdot))$. If $\hat{\mathbf{M}}$ is sufficiently accurate, then $\mathbb{E}[\|\mathbf{U}\mathbf{G}_t\mathbf{w}\|_2^2]\ll \mathbb{E}[\|\mathbf{G}_t\mathbf{w}\|_2^2]$.

After jamming suppression, the receiver performs OFDM demodulation, which includes signal segmentation over a length of $L_{\text{CP}}+L$, CP removal, and FFT operations, to obtain the OFDM observation grid $\tilde{\mathbf{Y}}\in\mathbb{C}^{L\times N_s}$. As these operations are all linear, $\tilde{\mathbf{Y}}$ can be written as
\begin{equation}
    \begin{aligned}
        \tilde{\mathbf{Y}}&=\text{vec}^{-1}\left(\mathbf{D}\mathbf{U}\mathbf{H}_t\mathbf{s}+\mathbf{D}\mathbf{U}\mathbf{G}_t\mathbf{w}+\mathbf{D}\mathbf{U}\mathbf{n}\right),
    \end{aligned}
\label{eq:ofdm_demod}
\end{equation}
where $\mathbf{D}\in\mathbb{C}^{LN_s\times N}$ is the matrix representation of OFDM demodulation, and $\text{vec}^{-1}:\mathbb{C}^{LN_s}\to\mathbb{C}^{L\times N_s}$ denotes the operation that partitions a vector into a matrix by filling columns sequentially.

Define $\mathbf{\Delta}:=\mathbf{I}_N-\mathbf{U}$. Since the cyclic prefix in a jamming-free scenario effectively mitigates inter-symbol interference (ISI), $\tilde{\mathbf{Y}}$ can be further expressed as
\begin{equation}
    \begin{aligned}
        \tilde{\mathbf{Y}}&=\underbrace{\mathbf{H}_f\odot\mathbf{S}}_{\text{Desired term}}+\underbrace{\text{vec}^{-1}\left(-\mathbf{D}\mathbf{\Delta}\mathbf{H}_t\mathbf{s}+\mathbf{D}\mathbf{U}\mathbf{G}_t\mathbf{w}+\mathbf{D}\mathbf{U}\mathbf{n}\right)}_{\text{Residual distortion term}},
    \end{aligned}
\label{eq:ofdm_demod_simp}
\end{equation}
where $\mathbf{S}\in\mathbb{C}^{L\times N_s}$ is the symbol grid obtained by stacking the frequency-domain symbols of all $LN_s$ subcarriers corresponding to $\mathbf{s}$ in columns, and $\mathbf{H}_f\in\mathbb{C}^{L\times N_s}$ denotes the CSI, i.e., the matrix representation of the channel frequency response (CFR) corresponding to $\mathbf{H}_t$. Unlike the multi-tap filter in the time domain represented by $\mathbf{H}_t$, $\mathbf{H}_f$ represents a single-tap filter for subcarriers in the frequency domain. It is important to note that the first term in \eqref{eq:ofdm_demod_simp} represents the desired signal in the absence of jamming, while the second term accounts for residual distortion, including suppression-induced distortion, residual jamming, and post-processing noise. These components significantly degrade signal quality.

The objective of joint channel estimation and data detection for jamming-resilient OFDM receivers is to obtain the optimal estimate $\mathbf{b}^{\ast}\in\{0,1\}^{N_b}$ of $\mathbf{b}$ by solving the following MAP optimization problem:
\begin{equation}
    \begin{aligned}
    \mathbf{b}^{\ast}&=\underset{\mathcal{\mathbf{b}}, \mathbf{H}_f}{\text{arg max}}\,\,\ln p\left(\mathbf{b},\mathbf{H}_f|\tilde{\mathbf{Y}}\right),
    \end{aligned}
\label{eq:map}
\end{equation}
where $p(\mathbf{b},\mathbf{H}_f|\tilde{\mathbf{Y}})$ denotes the posterior probability distribution of $\mathbf{b}$ and $\mathbf{H}_f$ given $\tilde{\mathbf{Y}}$. In practice, obtaining the optimal solution to \eqref{eq:map} is challenging. On the one hand, the analytical form of the posterior probability distribution is difficult to derive; on the other hand, the discrete nature of $\mathbf{b}$ complicates the optimization process. To address these challenges, we propose the BBD-JCED framework for jamming-resilient OFDM receivers to achieve more accurate bit recovery.

\section{Proposed Methodology}
\label{method}

This section details the proposed BBD-JCED framework, specifically designed to enhance joint channel estimation and data detection for jamming-resilient OFDM receivers. The methodology encompasses the overall architecture in subsection \ref{architecture}, followed by two key modules in subsections \ref{jsf} and \ref{bbjr}, and the corresponding training and inference algorithms in subsection \ref{algorithms}.

\subsection{Overall Architecture}
\label{architecture}

\begin{figure*}[!ht]
    \centering
    \includegraphics[width=0.99\textwidth]{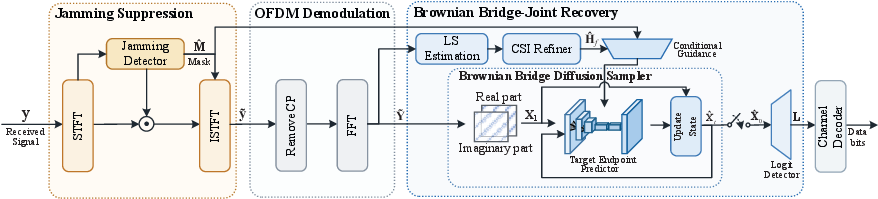}
    \caption{Overall architecture of the proposed BBD-JCED framework. The jamming suppression module first suppresses the dominant jamming components and produces the suppressed signal $\hat{\mathbf{y}}$ together with the spectral notching mask $\hat{\mathbf{M}}$. The OFDM demodulation module then converts $\hat{\mathbf{y}}$ into the OFDM observation grid $\tilde{\mathbf{Y}}$. Finally, the Brownian bridge-joint recovery module refines the CSI, performs conditional Brownian bridge diffusion sampling to recover the symbol grid, and generates LLRs for subsequent channel decoding.}
    \label{fig:architecture}
\end{figure*}

The overall architecture of the proposed BBD-JCED framework is illustrated in Fig. \ref{fig:architecture}. To improve the reliability of joint channel estimation and data detection in jamming-contested OFDM links, the proposed
framework adopts a multi-stage receiver architecture consisting of three modules:
\begin{enumerate}[label=(\roman*)]
\item \textbf{Jamming Suppression}: 
This module exploits the characteristics of the jamming signals in the STFT domain. A jamming detector first identifies the jamming-contaminated time-frequency bins and generates the spectral notching mask $\hat{\mathbf{M}}$. Spectral notching is subsequently applied to suppress the dominant jamming components, producing the suppressed time-domain signal $\tilde{\mathbf{y}}$ for subsequent receiver processing. Meanwhile, $\hat{\mathbf{M}}$ is retained as jamming-related side information for the downstream joint recovery module.
\item \textbf{OFDM Demodulation:} The suppressed time-domain signal $\tilde{\mathbf{y}}$ is transformed into the OFDM observation grid $\tilde{\mathbf{Y}}$ by CP removal and FFT processing.
\item \textbf{Brownian Bridge-Joint Recovery}: 
This module jointly refines the CSI and recovers the coded bits from the OFDM observation grid $\tilde{\mathbf{Y}}$. Specifically, a CSI refiner first improves the pilot-based LS estimate $\mathbf{H}_{\mathrm{LS}}$ to obtain the refined CSI $\hat{\mathbf{H}}_f$. The refined CSI and the spectral notching mask $\hat{\mathbf{M}}$ are then collected as physical side information to condition the subsequent data-recovery process. The real and imaginary components of $\hat{\mathbf{Y}}$ are stacked to form the terminal endpoint $\mathbf{X}_1$ of a Brownian bridge. Starting from $\mathbf{X}_1$, a conditional Brownian bridge diffusion sampler progressively reconstructs the clean symbol grid by repeatedly invoking the target-endpoint predictor under the conditional guidance. The resulting recovered target endpoint $\hat{\mathbf{X}}_0$ is subsequently mapped by the logit detector into bit-wise soft information, which is converted into the log-likelihood ratios (LLRs) required by the channel decoder.
\end{enumerate}

\subsection{Jamming Suppression}
\label{jsf}

The jamming suppression module is designed as a task-oriented front-end that reduces dominant jamming components before joint channel estimation and data detection. Instead of directly reconstructing the desired signal, the module formulates jamming suppression as a time-frequency masking problem \cite{ref:D.S}. This design is motivated by the fact that many types of narrowband and non-stationary jamming exhibit localized or structured occupancy in the time-frequency domain. The STFT representation therefore provides an explicit domain in which jamming-contaminated bins can be identified and selectively removed.

Specifically, let $\text{Cat}\left(\Re(\mathcal{F}(\mathbf{y})),\Im(\mathcal{F}(\mathbf{y}))\right)\in\mathbb{R}^{2\times K\times R}$ denote the two-channel real-valued representation of the received STFT coefficients, where $\text{Cat}(\cdot)$ denotes channel-wise concatenation. The trainable jamming detector first produces a jamming-probability map $\mathbf{P}\in [0,1]^{K\times R}$ given by
\begin{equation}
    \begin{aligned}
    \mathbf{P}=\mathcal{G}_{\theta}\left(\text{Cat}\left(\Re(\mathcal{F}(\mathbf{y})),\Im(\mathcal{F}(\mathbf{y}))\right)\right),
    \end{aligned}
\label{eq:jam_detect}
\end{equation}
where $\mathbf{P}_{k,r}$ represents the predicted probability that the $(k,r)$-th STFT bin is contaminated by jamming. $\mathcal{G}_{\theta}$ denotes the $\theta$-parametric jamming detector.

During training, let $\mathbf{M}\in\{0,1\}^{K\times R}$ denote the ground-truth notching mask, where $\mathbf{M}_{k,r}=0$ indicates a jamming-contaminated bin and $\mathbf{M}_{k,r}=1$ indicates an unjammed bin to be retained. Then, the jamming detector is optimized using the binary cross-entropy (BCE) loss given by
\begin{equation}
\begin{aligned}
\mathcal{L}_{\mathrm{JS}}
(\theta)&:=\mathbb{E}_{(\mathbf{y},\mathbf{M})}\left[-\frac{1}{KR}\langle\mathbf{1}-\mathbf{M},\ln\mathbf{P}\rangle_F \right. \\
&\left.\qquad\qquad\quad-\frac{1}{KR}\langle\mathbf{M},\ln(\mathbf{1}-\mathbf{P})\rangle_F\right].
\end{aligned}
\label{eq:js_loss}
\end{equation}
Minimizing \eqref{eq:js_loss} encourages $\mathbf{P}$ to approach one for jamming-contaminated bins and zero for unjammed bins, thereby providing a probabilistic basis for the subsequent binary notching decision.

During inference, the spectral notching mask $\hat{\mathbf{M}}$ is obtained by
\begin{equation}
    \begin{aligned}
    \hat{\mathbf{M}}_{k,r}=\mathbb{I}\left(\mathbf{P}_{k,r}<0.5\right),
    \end{aligned}
\label{eq:jam_detect_binary}
\end{equation}
Accordingly, $\hat{\mathbf{M}}_{k,r}=0$ indicates a detected jamming-contaminated bin, whereas $\hat{\mathbf{M}}_{k,r}=1$ indicates a retained bin. The resulting binary mask $\hat{\mathbf{M}}$ is then element-wise applied to the received STFT coefficients according to \eqref{eq:js}, yielding the suppressed signal $\tilde{\mathbf{y}}$.

The output of the jamming suppression module has two complementary roles in the proposed framework. First, $\tilde{\mathbf{y}}$ reduces the dominant jamming components presented to subsequent OFDM processing. Second, $\hat{\mathbf{M}}$ explicitly identifies the time-frequency bins removed due to detected jamming and is therefore retained as a conditioning variable for the Brownian bridge-joint recovery module. This side information informs the subsequent module which time-frequency bins have been removed by spectral notching, thereby preventing all elements of bins from being interpreted as uniformly reliable elements.

\begin{figure}[!ht]
    \centering
    \includegraphics[width=0.45\textwidth]{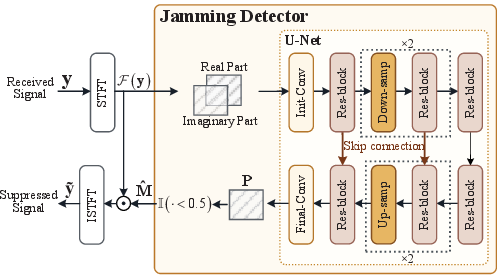}
    \caption{Structure of the jamming suppression module. The U-Net estimates a jamming-probability map $\mathbf{P}$ from the real and imaginary parts of the received STFT coefficients. During inference, a threshold operation converts $\mathbf{P}$ into the binary notching mask $\hat{\mathbf{M}}$, which is subsequently used for spectral notching.}
    \label{fig:jsf}
\end{figure}

The structure of the jamming suppression module is illustrated in Fig. \ref{fig:jsf}. Specifically, we adopt a U-Net-based jamming detector \cite{ref:O.R}, consisting of residual convolutional blocks together with downsampling and upsampling paths. The downsampling path enlarges the receptive field to capture structured jamming patterns across neighboring time-frequency bins, whereas the skip connections and upsampling path preserve the fine-grained localization information required for bin-wise jamming detection. Therefore, the U-Net serves as a suitable segmentation backbone for estimating the time-frequency jamming-probability map, while the resulting binary mask further provides explicit jamming-location information for the subsequent module.

\subsection{Brownian Bridge-Joint Recovery}
\label{bbjr}

As shown in \eqref{eq:ofdm_demod_simp}, although jamming suppression can eliminate dominant jamming components, the residual distortion terms in the OFDM observation grid $\tilde{\mathbf{Y}}$ still degrade signal quality for subsequent data recovery. To address this, we design the Brownian bridge-joint recovery module to perform joint channel estimation and data detection from $\tilde{\mathbf{Y}}$.

\subsubsection{Brownian Bridge-Based Probabilistic Graph}

\begin{figure}[!ht]
    \centering
    \includegraphics[width=0.45\textwidth]{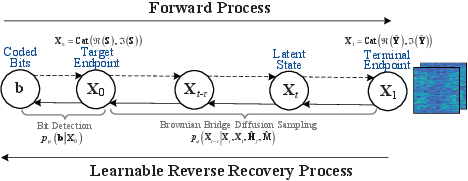}
    \caption{Brownian bridge-based probabilistic graph. The learnable reverse recovery process consists of two coupled stages, which are utilized to recover the coded bits $\mathbf{b}$ from the OFDM observation grid $\tilde{\mathbf{Y}}$.}
    \label{fig:prob_graph}
\end{figure}

Fig. \ref{fig:prob_graph} illustrates a Brownian bridge-based probabilistic graph used to model the probabilistic transition process from $\tilde{\mathbf{Y}}$ to the encoded bits $\mathbf{b}$. Specifically, the terminal endpoint $\mathbf{X}_1\in\mathbb{R}^{2\times (L-L_{\text{null}})\times N_s}$ of the Brownian bridge is set to the two-channel real-valued representation of the OFDM observation grid, i.e., $\text{Cat}(\Re(\tilde{\mathbf{Y}}),\Im(\tilde{\mathbf{Y}}))$, and the target endpoint $\mathbf{X}_0\in\mathbb{R}^{2\times (L-L_{\text{null}})\times N_s}$ is set to the two-channel real-valued  representation of the transmitted symbol grid, i.e., $\text{Cat}(\Re(\mathbf{S}),\Im(\mathbf{S}))$. Since $\mathbf{X}_0$ and $\mathbf{X}_1$ are resource-aligned representations of the transmitted symbol grid and its corresponding suppressed OFDM observation grid, a Brownian bridge is constructed to introduce a sequence of intermediate latent states $\{\mathbf{X}_t\in\mathbb{R}^{2\times (L-L_{\text{null}})\times N_s}\}_{t\in(0,1)}$ to define the transition path between them. Additionally, the transition path from the target endpoint $\mathbf{X}_0$ to the encoded bit $\mathbf{b}$ is also constructed in the probabilistic graph, corresponding to the bit detection process from the symbol information to the bit information. 

To incorporate the physical channel and jamming information into the recovery process, we further introduce a $\xi$-parametric CSI refiner $\mathcal{R}_{\xi}$ to refine the pilot-based LS estimate $\mathbf{H}_{\text{LS}}$ into a more accurate CSI estimate $\hat{\mathbf{H}}_f=\mathcal{R}_{\xi}(\mathbf{H}_{\text{LS}})\in\mathbb{C}^{L\times N_s}$. Then, $\hat{\mathbf{H}}_f$ and the spectral notching mask $\hat{\mathbf{M}}$ in \eqref{eq:jam_detect_binary} can be used as side information of the probabilistic graph, collectively denoted by $\mathbf{\Theta}=\{\hat{\mathbf{H}}_f, \hat{\mathbf{M}}\}$. Furthermore, the path from $\mathbf{X}_1$ to $\mathbf{b}$ can be characterized by a learnable reverse recovery process that consists of two coupled stages conditioned on this physical side information. Specifically, starting from the terminal endpoint $\mathbf{X}_1$, the Brownian bridge diffusion sampling progressively transforms the OFDM observation grid toward the transmitted symbol grid $\mathbf{X}_0$ with the $\phi$-parametric transition probability $p_{\phi}(\mathbf{X}_{t-\tau}|\mathbf{X}_t,\mathbf{X}_1,\mathbf{\Theta})$, after which the bit detection maps $\mathbf{X}_0$ to the bit-level posterior information with the $\psi$-parametric transition probability $p_{\psi}(\mathbf{b}|\mathbf{X}_0)$. In this manner, the refined CSI directly affects the recovery trajectory, thereby coupling channel estimation with data detection within a unified probabilistic framework.

\subsubsection{Variational Lower Bound and Training Objective}

Based on the probabilistic graph established above, the ultimate objective of the proposed BBD-JCED framework is to infer the coded bits $\mathbf{b}$ from the terminal endpoint $\mathbf{X}_1$, conditioned on $\mathbf{\Theta}=\{\hat{\mathbf{H}}_f, \hat{\mathbf{M}\}}$. The corresponding conditional bit posterior distribution is $p(\mathbf{b}|\mathbf{X}_1,\mathbf{\Theta})$. Directly maximizing this posterior distribution is generally intractable, but we can construct the training objective for the Brownian Bridge-joint recovery module based on its VLB.

To establish a finite-dimensional latent-variable representation, we discretize the bridge interval into
\begin{equation}
    \begin{aligned}
    0=t_0<t_1<\cdots<t_{\eta}<\cdots<t_T=1,
    \end{aligned}
\label{eq:t_interval}
\end{equation}
where $T\in\mathbb{N}$ is the number of intervals. The endpoint states satisfy $\mathbf{X}_{t_0}=\mathbf{X}_0$ and $\mathbf{X}_{t_T}=\mathbf{X}_1$.

Following the standard Brownian bridge formulation \cite{ref:B.L}, the distribution of $\mathbf{X}_{t_\eta}$ given $\mathbf{X}_{t_{\eta-1}}$ and $\mathbf{X}_{t_T}$ can be expressed as
\begin{equation}
    \begin{aligned}
    &p(\mathbf{X}_{t_{\eta}}|\mathbf{X}_{t_{\eta-1}},\mathbf{X}_{t_T}) \\
    =&\mathcal{N}\left(\mathbf{X}_{t_{\eta}};\frac{1-t_{\eta}}{1-t_{\eta-1}}\mathbf{X}_{t_{\eta-1}}+\frac{t_\eta-t_{\eta-1}}{1-t_{\eta-1}}\mathbf{X}_{t_T},\delta_{t_{\eta}|t_{\eta-1}}\mathbf{I}\right),
    \end{aligned}
\label{eq:transition_forward}
\end{equation}
with $\delta_{t_{\eta}|t_{\eta-1}}$ expressed as
\begin{equation}
    \begin{aligned}
    \delta_{t_{\eta}|t_{\eta-1}}=(1-t_\eta)-\frac{(1-t_\eta)^2}{1-t_{\eta-1}}.
    \end{aligned}
\label{eq:transition_forward_var}
\end{equation}
Then, the distribution of $\mathbf{X}_{t_\eta}$ given $\mathbf{X}_{t_0}$ and $\mathbf{X}_{t_T}$ can be obtained as
\begin{equation}
    \begin{aligned}
    &p(\mathbf{X}_{t_{\eta}}|\mathbf{X}_{t_0},\mathbf{X}_{t_T}) \\
    =&\mathcal{N}\left(\mathbf{X}_{t_{\eta}};(1-t_{\eta})\mathbf{X}_{t_0}+t_{\eta}\mathbf{X}_{t_T},t_{\eta}(1-t_{\eta})\mathbf{I}\right).
    \end{aligned}
\label{eq:transition_forward_state}
\end{equation}

Based on the main idea of denoising diffusion methods, we define the $\phi$-parametric distribution of $\mathbf{X}_{t_{\eta-1}}$ given $\mathbf{X}_{t_{\eta}}$, $\mathbf{X}_{t_T}$, and $\mathbf{\Theta}$ as
\begin{equation}
    \begin{aligned}
    &p_{\phi}(\mathbf{X}_{t_{\eta-1}}|\mathbf{X}_{t_{\eta}},\mathbf{X}_{t_T},\mathbf{\Theta})\\
    :=&\mathcal{N}\left(\mathbf{X}_{t_{\eta-1}};\boldsymbol{\mu}_{\phi}(\mathbf{X}_{t_\eta},t_\eta,\mathbf{\Theta});\delta_{t_{\eta-1}|t_\eta}\mathbf{I}\right),
    \end{aligned}
\label{eq:transition_reverse}
\end{equation}
where $\boldsymbol{\mu}_{\phi}(\mathbf{X}_{t_\eta},t_\eta,\mathbf{\Theta})$ is the $\phi$-parametric predicted mean value of this distribution, and $\delta_{t_{\eta-1}|t_\eta}$ is the preset variance at each timestep.

Using the chain rule of probability, the following decompositions hold:
\begin{equation}
    \begin{aligned}
    p(\mathbf{X}_{t_{0:T-1}}|\mathbf{b},\mathbf{X}_{t_T})&=p(\mathbf{X}_{t_{0}}|\mathbf{b})p(\mathbf{X}_{t_{1:T-1}}|\mathbf{X}_{t_{0}},\mathbf{X}_{t_T}),\\
    p(\mathbf{b},\mathbf{X}_{t_{0:T-1}}|\mathbf{X}_{t_T},\mathbf{\Theta})&=p_{\psi}(\mathbf{b}|\mathbf{X}_{t_0})p_{\phi}(\mathbf{X}_{t_{0:T-1}}|\mathbf{X}_{t_{T}},\mathbf{\Theta}).
    \label{eq:decomp}
\end{aligned}
\end{equation}

Furthermore, we extend the Markov-based modeling in \cite{ref:J.S}. Specifically, we assume that the forward process is modeled as a non-Markov process, while the reverse recovery process is modeled as a conditional Markov process. These properties can be expressed as
\begin{equation}
    \begin{aligned}
    p(\mathbf{X}_{t_{\eta}}|\mathbf{X}_{t_{0:\eta-1}},\mathbf{X}_{t_T})&=p(\mathbf{X}_{t_{\eta}}|\mathbf{X}_{t_{\eta-1}},\mathbf{X}_{t_0},\mathbf{X}_{t_T}),\\
    p_{\phi}(\mathbf{X}_{t_{\eta}-1}|\mathbf{X}_{t_{\eta:T}},\mathbf{\Theta})&=p_{\phi}(\mathbf{X}_{t_{\eta-1}}|\mathbf{X}_{t_{\eta}},\mathbf{X}_{t_T},\mathbf{\Theta}).
    \end{aligned}
\label{eq:transition_markov}
\end{equation}

\begin{theorem}\label{theorm:vlb}
Reformulating $\boldsymbol{\mu}_{\phi}(\mathbf{X}_{t_\eta},t_\eta,\mathbf{\Theta})$ in \eqref{eq:transition_reverse} as
\begin{equation}
    \begin{aligned}
    &\boldsymbol{\mu}_{\phi}(\mathbf{X}_{t_\eta},t_\eta,\mathbf{\Theta})\\
    =&\frac{t_{\eta-1}}{t_{\eta}}\mathbf{X}_{t_{\eta}}+\left(1-\frac{t_{\eta-1}}{t_{\eta}}\right)\mathcal{V}_{\phi}(\mathbf{X}_{t_\eta},\mathbf{X}_{t_T},\mathbf{\Theta},t_\eta),
    \end{aligned}
\label{eq:transition_reverse_mean}
\end{equation}
where $\mathcal{V}_{\phi}$ is the $\phi$-parametric target endpoint predictor. This name originates from the fact that the output $\mathbf{X}_{t_0|t_\eta}$ of $\mathcal{V}_{\phi}$ at timestep $t_{\eta}$ can be trained to approximate $\mathbf{X}_{t_0}$ in the subsequent discussion. Then, the VLB of $\ln p(\mathbf{b}|\mathbf{X}_{t_T},\mathbf{\Theta})$ can be expressed as
\begin{equation}
    \begin{aligned}
    &\ln p(\mathbf{b}|\mathbf{X}_{t_T},\mathbf{\Theta})\ge\text{VLB} \\
    \propto& -\sum_{\eta=1}^{T}C_{\eta}\mathbb{E}_{(\mathbf{X}_{t_0},\mathbf{X}_{t_T},\mathbf{\Theta})}\left[\left\|\mathcal{V}_{\phi}(\mathbf{X}_{t_\eta},\mathbf{X}_{t_T},\mathbf{\Theta},t_\eta)-\mathbf{X}_{t_0}\right\|_F^2\right]\\
    &+\mathbb{E}_{(\mathbf{b},\mathbf{X}_{t_0})}\left[\ln p_{\psi}(\mathbf{b}|\mathbf{X}_{t_0})\right],
    \label{eq:vlb_theorm}
\end{aligned}
\end{equation}
where $\{C_\eta\in\mathbb{R}_{>0}\}$ are constants independent of the learnable parameters $\phi$, $\psi$, and $\xi$.
\end{theorem}

\begin{proof}
The proof is relegated to Appendix \ref{appendix:vlb}.
\end{proof}

Theorem \ref{theorm:vlb} provides a variational interpretation for learning the reverse recovery process of the Brownian bridge-based probabilistic graph. The first term in \eqref{eq:vlb_theorm} optimizes the target endpoint predictor $\mathcal{V}_{\phi}$ toward the target endpoint $\mathbf{X}_{0}$, whereas the second term in \eqref{eq:vlb_theorm} maximizes the conditional log-likelihood of the coded bits given the clean endpoint. 

Motivated by \cite{ref:S.Z}, we assume a conditionally independent Bernoulli
distribution for $p_{\psi}\left(\mathbf{b}|\mathbf{X}_{t_0}\right)$, i.e.,
\begin{equation}
\begin{aligned}
p_{\psi}\left(\mathbf{b}|\mathbf{X}_{t_0}\right)=\prod_{i=1}^{N_b}\sigma\left(\mathcal{Z}_{\psi}(\mathbf{X}_{t_0})_{i}\right)^{\mathbf{b}_i}\left(1-\sigma(\mathcal{Z}_{\psi}(\mathbf{X}_{t_0})_{i})\right)^{1-\mathbf{b}_i},
\end{aligned}
\label{eq:bit_likelihood}
\end{equation}
where $\mathcal{Z}_{\psi}$ is the $\psi$-parametric logit detector, and $\sigma(\cdot)$ is the sigmoid function. Accordingly, the negative log-likelihood $-\ln p_\phi(\mathbf{b}|\mathbf{X}_{t_0})$ is exactly the BCE function, i.e.,
\begin{equation}
\begin{aligned}
-\ln p_\psi(\mathbf{b}|\mathbf{X}_{t_0})=&-\left\langle\mathbf{b},\ln \sigma(\mathcal{Z}_{\psi}(\mathbf{X}_{t_0}))\right\rangle_F\\
&-\left\langle1-\mathbf{b},1-\ln\sigma(\mathcal{Z}_{\psi}(\mathbf{X}_{t_0}))\right\rangle_F.
\end{aligned}
\label{eq:bit_bce}
\end{equation}

For practical training, we follow the commonly adopted simplified reweighting strategy \cite{ref:J.H} and replace the timestep-dependent coefficients $C_{\eta}$ by unity in \eqref{eq:vlb_theorm}. This yields the following endpoint-recovery loss:
\begin{equation}
\begin{aligned}
\mathcal{L}_{\text{BB}}(\phi,\xi):=\mathbb{E}_{(\mathbf{X}_0,\mathbf{X}_1,t,\mathbf{\Theta})}
\left[\left\|\mathcal{V}_{\phi}(\mathbf{X}_{t},\mathbf{X}_{1},\mathbf{\Theta},t)-\mathbf{X}_{0}\right\|_2^2\right].
\end{aligned}
\label{eq:bb_loss}
\end{equation}

On the other hand, the second term in \eqref{eq:vlb_theorm} is evaluated at the ground-truth target endpoint $\mathbf{X}_0$, while the logit detector $\mathcal{Z}_{\psi}$ operates on the recovered target endpoint $\hat{\mathbf{X}}_0$ during actual inference. Directly using the clean endpoint for bit-level training therefore introduces a mismatch between the training and finite-step sampling of the Brownian bridge. To alleviate this mismatch, we replace the clean endpoint bit-likelihood term by an inference-aligned surrogate evaluated on the output of the Brownian bridge diffusion sampler. Let
\begin{equation}
\hat{\mathbf{X}}_0=\mathcal{I}_{\phi}^{(J)}\left(\mathbf{X}_1,\mathbf{\Theta}\right),
\label{eq:sampler_operator}
\end{equation}
denote the target endpoint recovered by a $J$-step $\phi$-parametric Brownian bridge diffusion sampler $\mathcal{I}_{\phi}^{(J)}$ initialized from $\mathbf{X}_1$ (The detailed sampling procedure is presented in the following subsection). Accordingly, the inference-aligned bit detection loss is defined as
\begin{equation}
\begin{aligned}
\mathcal{L}_{\text{Logit}}(\phi,\psi,\xi):=\mathbb{E}_{(\mathbf{b},\mathbf{X}_1)}\left[
-\ln p_{\psi}\left(\mathbf{b}|\hat{\mathbf{X}}_0\right)\right],
\end{aligned}
\label{eq:inference_bit_loss}
\end{equation}
where the negative log-likelihood can also be calculated according to \eqref{eq:bit_bce}.

Since the refined CSI $\hat{\mathbf{H}}_f=\mathcal{R}_{\xi}(\mathbf{H}_{\text{LS}})$ is included in $\mathbf{\Theta}$, both
$\mathcal{L}_{\text{BB}}$ and $\mathcal{L}_{\text{Logit}}$ provide task-oriented supervision to the CSI refiner through the downstream recovery process. However, task-oriented optimization alone does not explicitly enforce the refined CSI to remain consistent with the underlying physical channel response. During supervised training, the ground-truth CSI $\mathbf{H}_f$ generated by the channel simulator is therefore utilized as an auxiliary physical supervision signal. Specifically, the CSI refinement loss is defined as
\begin{equation}
\mathcal{L}_{\mathrm{CSI}}(\xi):=\mathbb{E}\left[\left\|\hat{\mathbf{H}}_f-\mathbf{H}_f\right\|_F^2\right].
\label{eq:csi_refine_loss}
\end{equation}
The ground-truth CSI is required only during training and is not used during inference.

Combining \eqref{eq:bb_loss}, \eqref{eq:inference_bit_loss}, and \eqref{eq:csi_refine_loss}, the final VLB-guided joint training objective of the Brownian bridge-joint recovery module is given by
\begin{equation}
\begin{aligned}
\mathcal{L}_{\text{BB-JR}}(\phi,\psi,\xi):=&
\lambda_{\text{BB}}\mathcal{L}_{\text{BB}}(\phi,\xi)+\lambda_{\text{Logit}}\mathcal{L}_{\text{Logit}}(\phi,\psi,\xi)
\\
&+\lambda_{\text{CSI}}\mathcal{L}_{\text{CSI}}(\xi),
\end{aligned}
\label{eq:bb_jr_loss}
\end{equation}
where $\lambda_{\text{BB}}, \lambda_{\text{Logit}},\lambda_{\text{CSI}}\in\mathbb{R}_{>0}$ are balancing coefficients. The gradient of $\mathcal{L}_{\text{BB-JR}}(\phi,\psi,\xi)$ can propagate through the logit detector $\mathcal{Z}_{\psi}$, target endpoint predictor $\mathcal{V}_{\phi}$, and CSI refiner $\mathcal{R}_{\xi}$, thereby enabling comprehensive learning of the joint channel estimation and data detection task after jamming suppression.

\subsubsection{Diffusion Sampling and Network Backbone}

We next specify the Brownian bridge diffusion sampler $\mathcal{I}_{\phi}^{(J)}$ introduced in \eqref{eq:sampler_operator} and describe the neural-network implementation of the Brownian bridge-joint recovery module.

To reduce the inference complexity, instead of traversing all the discretized bridge timesteps, we uniformly select timesteps over $[0,1]$ with a constant sampling interval $1/J$, i.e.,
\begin{equation}
\mathcal{T}=\{\tau_j\}_{j=0}^{J},\qquad \tau_j=\frac{j}{J},
\label{eq:sampling_subsequence}
\end{equation}
where $\mathcal{T}$ is the sampling sequence. Then, the sampler is initialized from the terminal endpoint as
\begin{equation}
\hat{\mathbf{X}}_{\tau_J}=\mathbf{X}_1.
\label{eq:sampling_init}
\end{equation}

At the $j$-th sampling step, the output of the target endpoint predictor $\mathcal{V}_{\phi}$ is used to compute \eqref{eq:transition_reverse_mean}, and update the next state $\hat{\mathbf{X}}_{\tau_{j-1}}$ according to \eqref{eq:transition_reverse} as follows:
\begin{equation}
\begin{aligned}
\hat{\mathbf{X}}_{\tau_{j-1}}
=&\frac{\tau_{j-1}}{\tau_j}\hat{\mathbf{X}}_{\tau_j}+\left(1-\frac{\tau_{j-1}}{\tau_j}\right)\mathcal{V}_{\phi}(\hat{\mathbf{X}}_{\tau_{j}},\mathbf{X}_{1},\mathbf{\Theta},\tau_j)\\
&+\sqrt{\delta_{\tau_{j-1}|\tau_j}}\boldsymbol{\varepsilon}_j,
\end{aligned}
\label{eq:bb_sampling}
\end{equation}
where $\boldsymbol{\varepsilon}_j\sim\mathcal{N}(\mathbf{0},\mathbf{I})$. For low-complexity and reproducible receiver inference, we employ a deterministic sampling rule \cite{ref:J.S} by setting $\delta_{\tau_{j-1}|\tau_j}=0$. Accordingly,
\eqref{eq:bb_sampling} reduces to
\begin{equation}
\begin{aligned}
\hat{\mathbf{X}}_{\tau_{j-1}}
=\frac{\tau_{j-1}}{\tau_j}\hat{\mathbf{X}}_{\tau_j}+\left(1-\frac{\tau_{j-1}}{\tau_j}\right)\mathcal{V}_{\phi}(\hat{\mathbf{X}}_{\tau_{j}},\mathbf{X}_{1},\mathbf{\Theta},\tau_j).
\end{aligned}
\label{eq:deterministic_bb_sampling}
\end{equation}
Recursively applying \eqref{eq:deterministic_bb_sampling} from $\tau_J$ to $\tau_0$ defines the $J$-step $\phi$-parametric Brownian bridge diffusion sampler $\mathcal{I}_{\phi}^{(J)}$, which is exactly the recovered target endpoint employed in \eqref{eq:sampler_operator}.

\begin{figure*}[!ht]
    \centering
    \includegraphics[width=0.78\textwidth]{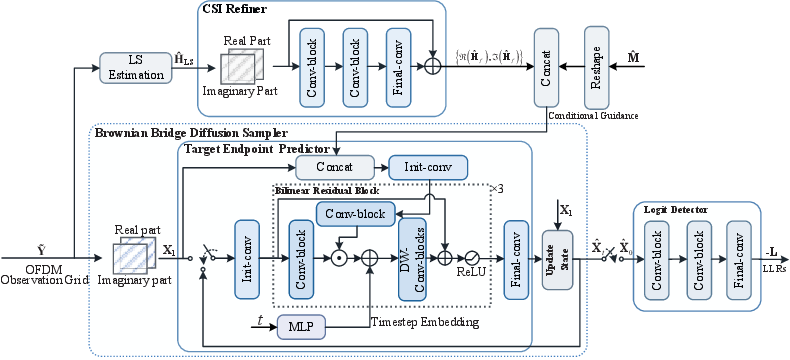}
    \caption{Structure of the Brownian bridge-joint recovery module. The CSI refiner improves the LS estimate and combines the refined CSI with the spectral notching mask to provide conditional guidance. The target-endpoint predictor is iteratively invoked by the Brownian bridge diffusion sampler to recover the target endpoint, which is subsequently mapped by the logit detector into soft LLRs for channel decoding.}
    \label{fig:bb_jr}
\end{figure*}

Based on the above discussion, the Brownian bridge-joint recovery module contains three learnable components: the CSI refiner $\mathcal{R}_{\xi}$, the target-endpoint predictor $\mathcal{V}_{\phi}$, and the logit detector $\mathcal{Z}_{\psi}$, as illustrated in Fig. \ref{fig:bb_jr}. Specifically, the CSI refiner processes the real and imaginary parts of the LS estimate $\mathbf{H}_{\text{LS}}$ through convolutional blocks with a residual connection to obtain the refined CSI $\hat{\mathbf{H}}_f$. The two-channel real-valued representation of $\hat{\mathbf{H}}_f$ is then concatenated with the spectral notching mask $\hat{\mathbf{M}}$ and transformed into conditional guidance for the reverse recovery process. For the target-endpoint predictor, the real and imaginary parts of the current bridge state are successively processed by an initial convolution, three bilinear residual blocks \cite{ref:X.M}, and a final convolution, where the timestep embedding and conditional guidance are injected into the intermediate features. The predictor is repeatedly invoked according to \eqref{eq:deterministic_bb_sampling} to progressively recover the target endpoint $\hat{\mathbf{X}}_0$. Finally, the logit detector maps $\hat{\mathbf{X}}_0$ into bit-wise logits $\hat{\mathbf{z}}$. Following the log-probability-ratio-based soft demodulation principle in \cite{ref:S.Z}, when $\sigma(\hat{\mathbf{z}}_i)$ represents the posterior probability of $\mathbf{b}_i=1$, the $i$-th LLR defined as $\mathbf{L}_i=\ln\frac{p(\mathbf{b}_i=0|\hat{\mathbf{X}}_0)}{p(\mathbf{b}_i=1|\hat{\mathbf{X}}_0)}$ is directly obtained as $-\hat{\mathbf{z}}_i$, which is subsequently fed to the channel decoder.

\subsection{Training and Inference Algorithms}
\label{algorithms}

Based on the preceding module designs, we summarize the training and inference procedures of the proposed BBD-JCED framework. 

During training, the jamming suppression module and the Brownian bridge-joint recovery module are optimized according to their corresponding objectives. For the jamming suppression module, the jamming detector is trained using the ground-truth spectral mask to distinguish jamming-contaminated and reliable time-frequency bins. The resulting detector is then used to generate the spectral notching mask $\hat{\mathbf{M}}$ for the subsequent receiver processing. For the Brownian bridge-joint recovery module, the transmitted symbol grid and the suppressed OFDM observation grid are first constructed as the paired bridge endpoints $\mathbf{X}_0$ and $\mathbf{X}_1$, respectively. The CSI refiner $\mathcal{R}_{\xi}$ produces the refined CSI $\hat{\mathbf{H}}_f$, which, together with $\hat{\mathbf{M}}$, forms the physical condition $\mathbf{\Theta}$. A bridge timestep $t$ is then sampled to generate the intermediate latent state $\mathbf{X}_t$, from which the target-endpoint predictor $\mathcal{V}_{\phi}$ learns to recover $\mathbf{X}_0$. Meanwhile, the $J$-step sampler $\mathcal{I}_{\phi}^{(J)}$ reconstructs the finite-step endpoint $\hat{\mathbf{X}}_0$ from $\mathbf{X}_1$, and the logit detector $\mathcal{Z}_{\psi}$ maps $\hat{\mathbf{X}}_0$ into bit-wise soft information. The CSI refiner, target-endpoint predictor, and logit detector are jointly optimized using the training objective defined in \eqref{eq:bb_loss}, such that the refined CSI is constrained by both physical channel supervision and the downstream data recovery. The complete training procedure is summarized in Algorithm \ref{algo:training}.

During inference, no ground-truth CSI, transmitted symbols, or coded bits are required. Given the received signal $\mathbf{y}$, the jamming suppression module first estimates $\hat{\mathbf{M}}$ and performs spectral notching, after which the OFDM demodulation module generates the observation grid $\tilde{\mathbf{Y}}$. The CSI refiner then refine the pilot-based LS estimate $\mathbf{H}_{\text{LS}}$ to $\hat{\mathbf{H}}_f$ and constructs $\mathbf{\Theta}=\{\hat{\mathbf{H}}_f,\hat{\mathbf{M}}\}$. The Brownian bridge diffusion sampler is initialized with the terminal endpoint $\hat{\mathbf{X}}{\tau_J}=\mathbf{X}_1$ and recursively applies \eqref{eq:deterministic_bb_sampling} over the sampling subsequence $\mathcal{T}$ to obtain $\hat{\mathbf{X}}_0$. Finally, the logit detector produces the bit-wise logits, which are converted into LLRs and delivered to the channel decoder. The overall inference procedure is summarized in Algorithm \ref{algo:inference}.

\begin{algorithm}[!h]
\caption{Training Procedure of BBD-JCED}
\label{algo:training}
\KwIn{Training dataset $\mathcal{D}=\{(\mathbf{y},\mathbf{M},\mathbf{b},\mathbf{S},\mathbf{H}_f\})$, jamming detector $\mathcal{G}_{\theta}$, CSI refiner $\mathcal{R}_{\xi}$, target-endpoint predictor $\mathcal{V}_{\phi}$, logit detector $\mathcal{Z}_{\psi}$, and sampling steps $J$.}

\small{
\Repeat{\textup{\textbf{converged}}}{
Sample a mini-batch from $\mathcal{D}$\;
Estimate the jamming-probability map $\mathbf{P}$ via \eqref{eq:jam_detect}\;
Update $\theta$ by minimizing \eqref{eq:js_loss}\;
}
Freeze $\theta$\;
\Repeat{\textup{\textbf{converged}}}{
Sample a mini-batch from $\mathcal{D}$\;
Estimate the jamming-probability map $\mathbf{P}$ via \eqref{eq:jam_detect}\;
Obtain the spectral notching mask $\hat{\mathbf{M}}$ via \eqref{eq:jam_detect_binary}\;
Obtain the suppressed signal $\tilde{\mathbf{y}}$ via \eqref{eq:js}\;
Obtain the OFDM observation grid $\tilde{\mathbf{Y}}$ via \eqref{eq:ofdm_demod}\;
Calculate the pilot-based LS estimate $\hat{\mathbf{H}}_{\text{LS}}$\;
Estimate the refined CSI $\hat{\mathbf{H}}_f=\mathcal{R}_{\xi}(\hat{\mathbf{H}}_{\text{LS}})$\;
Construct the terminal endpoint $\mathbf{X}_1=\text{Cat}(\Re(\tilde{\mathbf{Y}}),\Im(\tilde{\mathbf{Y}}))$ and the target endpoint $\mathbf{X}_0=\text{Cat}(\Re(\mathbf{S}),\Im(\mathbf{S}))$\;
Construct the condition $\mathbf{\Theta}=\{\hat{\mathbf{H}}_f,\hat{\mathbf{M}}\}$\;
Sample the timestep $t\in(0, 1)$\;
Sample the latent state $\mathbf{X}_{t}$ via \eqref{eq:transition_forward_state}\;
Predict the target endpoint $\mathbf{X}_{0|t}=\mathcal{V}_{\phi}(\mathbf{X}_t,\mathbf{X}_1,\mathbf{\Theta},t)$\;
Construct the uniform sampling sequence $\mathcal{T}$ via \eqref{eq:sampling_subsequence}\;
Recover $\hat{\mathbf{X}}_0$ via recursively applying \eqref{eq:deterministic_bb_sampling}\;
Calculate the detected logits $\mathcal{Z}_{\psi}(\hat{\mathbf{X}}_0)$\;
Update $\phi$, $\psi$, and $\xi$ by minimizing \eqref{eq:bb_jr_loss}\;
}
}
\KwOut{Optimized parameters $\theta$, $\phi$, $\psi$, and $\xi$.}
\end{algorithm}

\begin{algorithm}
\caption{Inference Procedure of BBD-JCED}
\label{algo:inference}
\KwIn{Received signal $\mathbf{y}$, trained jamming detector $\mathcal{G}_{\theta}$, trained CSI refiner $\mathcal{R}_{\xi}$, trained target-endpoint predictor $\mathcal{V}_{\phi}$, trained logit detector $\mathcal{Z}_{\psi}$, and sampling steps $J$.}
\small{
Estimate the jamming-probability map $\mathbf{P}$ via \eqref{eq:jam_detect}\;
Obtain the spectral notching mask $\hat{\mathbf{M}}$ via \eqref{eq:jam_detect_binary}\;
Obtain the suppressed signal $\tilde{\mathbf{y}}$ via \eqref{eq:js}\;
Obtain the OFDM observation grid $\tilde{\mathbf{Y}}$ via \eqref{eq:ofdm_demod}\;
Calculate the pilot-based LS estimate $\hat{\mathbf{H}}_{\text{LS}}$\;
Estimate the refined CSI $\hat{\mathbf{H}}_f=\mathcal{R}_{\xi}(\hat{\mathbf{H}}_{\text{LS}})$\;
Construct the terminal endpoint $\mathbf{X}_1=\text{Cat}(\Re(\tilde{\mathbf{Y}}),\Im(\tilde{\mathbf{Y}}))$\;
Construct the condition $\mathbf{\Theta}=\{\hat{\mathbf{H}}_f,\hat{\mathbf{M}}\}$\;
Construct the uniform sampling sequence $\mathcal{T}$ via \eqref{eq:sampling_subsequence}\;
Recover $\hat{\mathbf{X}}_0$ via recursively applying \eqref{eq:deterministic_bb_sampling}\;
Calculate the detected logits $\mathbf{z}=\mathcal{Z}_{\psi}(\hat{\mathbf{X}}_0)$\;
Calculate the LLRs $\mathbf{L}=-\mathbf{z}$\;
Perform channel decoding using $\mathbf{L}$\;
}
\KwOut{Decoded data bits.}
\end{algorithm}

\section{Simulation Results}
\label{result}

In this section, we validate the effectiveness of the proposed BBD-JCED framework. Comprehensive simulations are conducted to benchmark the BBD-JCED framework against representative existing joint channel estimation and data detection schemes under identical communication and jamming condition.

\subsection{Simulation Setup}

\subsubsection{System Parameters}
\label{system_para}

Based on the system model in Section \ref{model}, the parameter configuration of the OFDM communication link is summarized in Table \ref{table:para_ofdm}. The OFDM link operates at $2.1$ GHz (sub-$6$ GHz band) with a subcarrier spacing of $\Delta f=30$ kHz. The OFDM symbol indices carrying pilot symbols are configured as $(2,5,8,11)$ according to the PDSCH DM-RS positions in Table $7.4.1.1.2$-$3$ of TS $38.211$ \cite{ref:3gpp}. For channel modeling, we consider the tapped delay line (TDL) models specified in TS $38.901$ \cite{ref:3gpp-2}. The delay spread is set to $100$ ns, and the maximum Doppler shift is set to $700$ Hz.

\begin{table}[!ht]
\caption{Parameters of the OFDM communication link}
\label{table:para_ofdm}
\centering
\scalebox{0.9}{
\begin{tabular}{|c|c|}
\hline
\textbf{Parameter} &\textbf{Value} \\
\hline
FFT size  &$L=256$ \\
\hline
CP length  &$L_{\text{CP}}=18$ \\
\hline
OFDM symbols per slot  &$N_s=14$ \\
\hline
Guard subcarriers per OFDM symbol  &$L_{\text{null}}=16 (\text{Outer})+2 (\text{DC})$  \\
\hline
Subcarriers spacing  &$\Delta f=30$ kHz \\
\hline
\multirow{2}*{Pilot symbols}  &DM-RS positions:$(2,5,8,11)$ \\
                       &with constant value $e^{j\pi/4}$  \\
\hline                     
Constellation modulation  &QPSK \\
\hline                     
Channel code  &Rate-$0.2$ LDPC code \\
\hline
Center frequency  &$2.1$ GHz \\
\hline
Noise level  &$\text{SNR}\in[0, 40]$ dB \\
\hline
Channel model  &TDL models\\
\hline
\end{tabular}
}
\end{table}

The parameter configuration of the jamming link is summarized in Table \ref{table:para_jam}. We consider two representative jamming types: CSN jamming and LFM jamming. CSN jamming is assumed to contain $I\in\{1,\cdots,L-L_{\text{null}}\}$ combs within the OFDM bandwidth, where each comb occupies with bandwidth $\Delta f$. LFM jamming is assumed to have $Z\in\mathbb{R}_{>0}$ periods per OFDM slot (or per $N_s$ OFDM symbols). Since the jamming link typically involves a high-power LOS path \cite{ref:P.W-2}, we adopt a flat Rician model with a K-factor of $15$ dB and a maximum Doppler shift of $70$ Hz for the jamming channel.

\begin{table}[!h]
\caption{Parameters of the jamming link}
\label{table:para_jam}
\centering
\scalebox{0.9}{
\begin{tabular}{|c|c|}
\hline
\textbf{Parameter} &\textbf{Value} \\
\hline
Jamming type  &CSN or LFM jamming \\
\hline
\multirow{2}*{Jamming parameters} & CSN: $I$ combs within the OFDM bandwidth\\ \cline{2-2} 
                                  & LFM: $Z$ periods per slot\\
\hline
Jamming intensity  &$\text{SJR}\in[-50, 0]$ dB \\
\hline
Center frequency  &$2.1$ GHz \\
\hline
Channel model  &Flat Rician model \\
\hline
\end{tabular}
}
\end{table}

\subsubsection{Model Hyperparameters}

Based on the methodology in Section \ref{method}, the hyperparameter configuration of the proposed BBD-JCED framework is summarized in Table \ref{table:para_model}. The dataset includes samples under CSN or LFM jamming, where the signal-to-jamming ratio (SJR) and the signal-to-noise ratio (SNR) are uniformly sampled based on Table \ref{table:para_ofdm} and \ref{table:para_jam}. For the network structure, the convolutional layer is represented as Conv(channel, kernel, stride, padding), where channel, kernel, stride, and padding refer to the channel dimensions, convolutional kernel size, stride size, and padding size, respectively. The depthwise convolutional layer is represented as DW-Conv(channel, kernel, stride, padding). ReLU denotes the rectified linear unit. Pool($2$) denotes the $2$-stride average pooling layer and Interp($2$) denotes the nearest-neighbor interpolation layer with a scale factor of $2$. Additionally, the values of all loss weights are determined by the validation set.

\begin{table}[!h]
\caption{Hyperparameters of the proposed BBD-JCED Framework}
\label{table:para_model}
\centering
\scalebox{0.9}{
\begin{tabular}{|c|c|}
\hline
\textbf{Parameter} &\textbf{Value} \\
\hline
\multirow{2}*{STFT processing}  &Frequency-domain bins $K=256$ \\
                                &and time-domain bins $R=31$ \\
\hline
\multirow{4}*{\makecell{Network structure of the \\ jamming detector}}  &Channel list: $c\in\{2, 4, 8\}$ \\
                                            &Res-block: $2\times$ \{Conv($c$, $3$, $1$, $1$)+ReLU\}\\
                                            &Down-samp: Conv($c$, $3$, $2$, $1$)+Pool($2$)  \\
                                            &Up-samp: Conv($c$, $3$, $1$, $1$)+Interp($2$)  \\
\hline
\multirow{2}*{\makecell{Network structure of the \\ CSI Refiner}}  &Res-block: $2\times$ \{Conv($32$, $3$, $1$, $1$)+ReLU\}\\
&Final-conv: Conv($2$, $3$, $1$, $1$)\\
\hline
\multirow{2}*{\makecell{Network structure of the \\ logit detector}}  &$2\times$ \{Conv($16$, $3$, $1$, $1$)+ReLU\} \\
&+Conv($2$, $1$, $1$, $0$)\\
\hline
\multirow{2}*{\makecell{Network structure of the \\ target endpoint predictor}}  &Bilinear residual block: $2\times$Conv($32$, $1$, $1$, $0$)+ \\
                            &+$2\times$DW-Conv($32$, $3$, $1$, $1$)+ReLU \\
\hline
Sampling steps  &$J=3$  \\
\hline
\multirow{2}*{Dataset size}  &A total of $20000$ samples, \\
                              &and $\text{training:validation:test}=8\text{:}1\text{:}1$ \\ 
\hline
\makecell{Epochs for the jamming \\suppression module}  &200 \\
\hline
\makecell{Epochs for the Brownian \\bridge-joint recovery module}  &600 \\
\hline
Batch size &64 \\
\hline
Optimizer  &AdamW with a learning rate of $10^{-4}$  \\
\hline
Loss weights  &$\lambda_\text{BB}=0.25$, $\lambda_{\text{Logit}}=1$, and $\lambda_{\text{CSI}}=0.05$  \\
\hline
\end{tabular}
}
\end{table}

\subsection{Performance Evaluation}

To rigorously evaluate the proposed BBD-JCED framework, we first assess the jamming suppression module and then evaluate the overall framework in terms of bit error rate (BER).

\subsubsection{Jamming Suppression Performance}

To quantitatively evaluate the jamming suppression performance, we adopt the scale-invariant source-to-noise ratio (SI-SNR) metric \cite{ref:Y.L-2}, given by
\begin{equation}
    \begin{aligned}
        \text{SI-SNR}(\tilde{\mathbf{y}})&=\mathbb{E}\left[10\lg\frac{\left\|(\mathbf{H}_t\mathbf{s})^{\mathrm{H}}\cdot\tilde{\mathbf{y}}\cdot\mathbf{H}_t\mathbf{s}\big/\|\mathbf{H}_t\mathbf{s}\|_2^2\right\|_2^2}{\left\|\tilde{\mathbf{y}}-(\mathbf{H}_t\mathbf{s})^{\mathrm{H}}\cdot\tilde{\mathbf{y}}\cdot\mathbf{H}_t\mathbf{s}\big/\|\mathbf{H}_t\mathbf{s}\|_2^2\right\|_2^2}\right],
    \end{aligned}
\label{eq:nmse}
\end{equation}
where SI-SNR is measured in dB, and all variables follow the definitions in \eqref{eq:rx_signal} and \eqref{eq:js}. A higher SI-SNR indicates better suppression quality and a higher effective SJNR.

We compare the proposed module with two representative anti-jamming baselines and the case without suppression. \emph{No Suppression} directly uses the received signal $\mathbf{y}$. \emph{WaveNet} is a time-domain data-driven signal separation baseline \cite{ref:A.L} that exploits stacked dilated convolutions to capture long-range temporal structures. \emph{TF-Unet} is a learning-based time-frequency jamming mitigation baseline \cite{ref:J.S-v2} that employs a U-Net to reconstruct the jamming-free STFT spectrum.

For a fair comparison,  and the proposed module employ the same STFT configuration unless otherwise specified. Tables \ref{table:si_snr_csn} and \ref{table:si_snr_lfm} report the SI-SNR results under CSN and LFM jamming, respectively. The TDL-A channel is considered for the OFDM communication link, the SNR is fixed at $20$ dB, the SJR varies from $-40$ dB to $-20$ dB in $5$ dB increments. The combs of CSN jamming is set to $I=40$, and the sweep periods of LFM jamming is set to $Z=6$. The proposed jamming suppression module consistently achieves the highest SI-SNR across all evaluated SJR values. This performance advantage can be attributed to its explicit exploitation of the time-frequency characteristics of jamming, whereby the learned mask enables selective spectral notching of contaminated time-frequency components. Compared with the time-domain signal separation network \emph{WaveNet} and time-frequency-domain spectrum reconstruction network \emph{TF-Unet}, the proposed module is particularly effective under severe jamming conditions.

The inference complexity of all jamming suprression methods is compared in Table \ref{table:complexity_js}. The input dimension is denoted as $(B,C,D_1,D_2)$, where $B$ is the batch size, $C$ denotes the real and imaginary channels, and $D_1$ and $D_2$ are the frequency- and time-domain dimensions, respectively. For \emph{WaveNet}, $D_2=N_s(L+L_{\text{CP}})=3836$ from Table \ref{table:para_ofdm}, whereas for \emph{TF-Unet} and the proposed module, $D_1$ and $D_2$ correspond to $K$ and $R$ in Table \ref{table:para_model}. With comparable parameter counts, the proposed module requires $33.3\times10^{6}$ floating-point operations (FLOPs), compared with $97.2\times10^{6}$ and $209.0\times10^{6}$ FLOPs for \emph{WaveNet} and \emph{TF-Unet}, corresponding to $65.7\%$ and $84.1\%$ reductions, respectively. This demonstrates that the proposed module achieves superior jamming suppression performance while maintaining a relatively low inference cost.

\begin{table}[!h]
\caption{SI-SNR ($\textnormal{dB}$) results under CSN jamming with $I=40$}
\label{table:si_snr_csn}
\centering
\scalebox{0.9}{
\begin{tabular}{c|ccccc}
\toprule
\multirow{2}*{\textbf{Method}}
& \multicolumn{5}{c}{\textbf{SJR (dB)}} \\
& -40 & -35 & -30 & -25 & -20 \\
\midrule
No Suppression
& -36.31 & -34.10 & -30.17 & -25.22 & -20.19 \\
WaveNet
& -34.90 & -30.10 & -25.46 & -19.86 & -11.26 \\
TF-Unet
& -28.19 & -22.68 & -15.42 & -7.38 & -2.41 \\
Proposed Module
& \textbf{-19.66} & \textbf{-14.59} & \textbf{-9.86}
& \textbf{-5.49} & \textbf{-2.10} \\
\bottomrule
\end{tabular}
}
\end{table}

\begin{table}[!h]
\caption{SI-SNR ($\textnormal{dB}$) results under LFM jamming with $Z=6$}
\label{table:si_snr_lfm}
\centering
\scalebox{0.9}{
\begin{tabular}{c|ccccc}
\toprule
\multirow{2}*{\textbf{Method}}
& \multicolumn{5}{c}{\textbf{SJR (dB)}} \\
& -40 & -35 & -30 & -25 & -20 \\
\midrule
No Suppression
& -36.52 & -34.81 & -30.06 & -25.01 & -20.17 \\
WaveNet
& -33.15 & -28.66 & -22.08 & -14.75 & -8.02 \\
TF-Unet
& -17.73 & -12.60 & -7.01 & -1.45 & 1.36 \\
Proposed Module
& \textbf{-9.92} & \textbf{-5.42} & \textbf{-1.70}
& \textbf{0.73} & \textbf{1.91} \\
\bottomrule
\end{tabular}
}
\end{table}

\begin{table}[!ht]
\caption{Inference complexity of tested jamming suppression methods}
\label{table:complexity_js}
\centering
\scalebox{0.9}{
\begin{tabular}{c|c|c|c}
\toprule
\textbf{Method} &\textbf{\makecell{Input Dimension \\ $(B,C,D_1,D_2)$}} &\makecell{\textbf{FLOPs} \\($\times 10^{6}$)} &\makecell{\textbf{Parameters}\\ ($\times 10^4$)}\\
\midrule
WaveNet &(1,2,$\emptyset$,3836) &97.2 &1.4 \\
TF-Unet &\multirow{2}*{(1,2,256,31)}  &209.0 &1.4 \\
Proposed Module & &\textbf{33.3} &1.4  \\
\bottomrule
\end{tabular}
}
\end{table}

\subsubsection{Bit Recovery Performance}

Building on the improved SJNR provided by the jamming suppression module, we evaluate the overall BBD-JCED framework in terms of BER. Four receiver baselines are considered. \emph{TraditionRx} performs pilot-based LS estimation, linear interpolation, MMSE equalization, and minimum-Euclidean-distance demodulation. \emph{DECNN} is a CNN-based receiver adapted from \cite{ref:J.L}, which directly maps the suppressed OFDM observation grid and pilot-based LS estimate to detected data symbols through a single feed-forward mapping. \emph{DBRNet} is a diffusion-bridge-assisted joint estimation and detection receiver adapted from \cite{ref:R.Z}, where data symbols and CSI are iteratively recovered through multi-step bridge-diffusion sampling. \emph{DM-JCED} is a standard diffusion-assisted counterpart based on \cite{ref:H.S-TVT}, in which the Brownian bridge is replaced by a standard diffusion process; the suppressed OFDM observation grid serves as conditioning information rather than the terminal endpoint, and clean data symbols are progressively recovered through multi-step denoising.

For a fair comparison of the generative receivers, \emph{DBRNet}, \emph{DM-JCED}, and BBD-JCED all use $3$ sampling steps. Moreover, all baseline receivers employ the proposed jamming suppression module as a common front-end to isolate differences in their back-end joint recovery capabilities. The data BER is defined as
\begin{equation}
\begin{aligned}
\text{Data BER}&=\mathbb{E}\left[\mathbb{I}\left(\hat{\mathbf{b}}_{c}\neq \mathbf{b}_{c}\right)\right],
\end{aligned}
\label{eq_ber_fec}
\end{equation}
where $\mathbf{b}_c$ and $\hat{\mathbf{b}}_c$ denote the transmitted and decoded data bits, respectively.

\begin{figure*}[ht]
    \centering
    \subfigure[$I=32$]{
        \centering
        \includegraphics[width=0.39\textwidth]{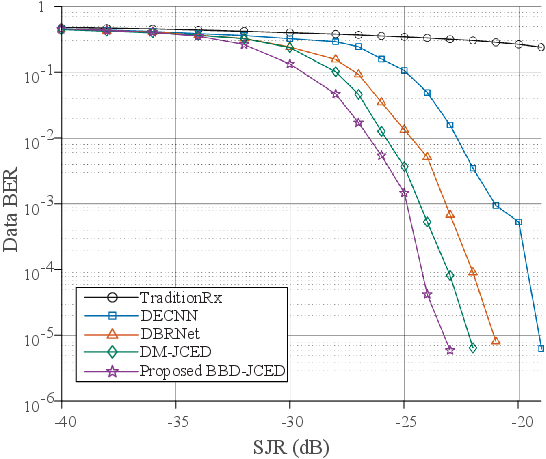}
    }
    \subfigure[$I=40$]{
        \centering
        \includegraphics[width=0.39\textwidth]{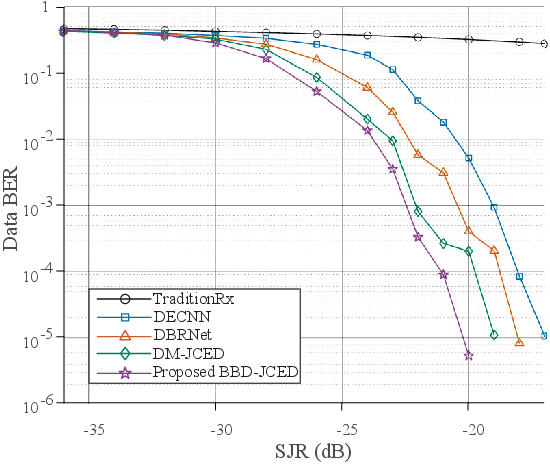}
    }
    \caption{Data BER performance under CSN jamming with different numbers of combs $I$.}
\label{fig:eval_csn}
\end{figure*}

Fig. \ref{fig:eval_csn} shows the data BER under CSN jamming with different numbers of combs $I$. The TDL-A channel model is considered for the OFDM communication link and the SNR is fixed at $20$ dB. TraditionRx exhibits a pronounced error floor under both $I=32$ and $I=40$, indicating that conventional LS estimation and MMSE equalization remain sensitive to the residual distortion after jamming suppression. The learning-based receivers substantially improve the BER performance, with the generative \emph{DBRNet} and \emph{DM-JCED} generally outperforming the single-step discriminative \emph{DECNN}. Among all compared schemes, BBD-JCED consistently achieves the lowest BER over the evaluated SJR range. In particular, at a target BER of approximately $10^{-5}$, BBD-JCED tolerates an SJR about $1\sim5$ dB lower than the considered learning-based baselines, depending on $I$. When $I$ increases from $32$ to $40$, the BER curves of all schemes shift toward higher SJR values because a larger portion of the OFDM spectrum is affected by jamming; nevertheless, the proposed framework maintains its performance advantage, demonstrating its robustness to variations in the CSN jamming parameters.

\begin{figure*}[ht]
    \centering
    \subfigure[$Z=4.5$]{
        \centering
        \includegraphics[width=0.39\textwidth]{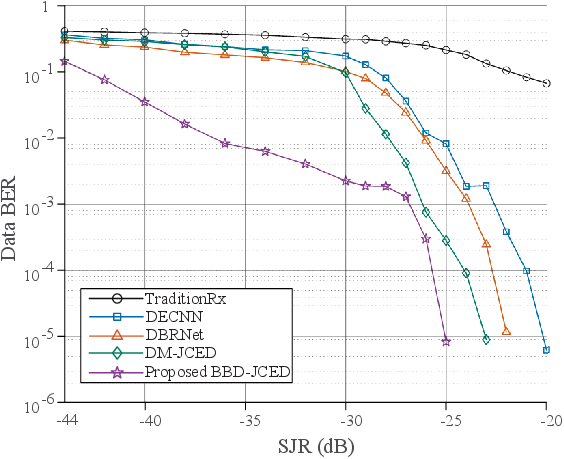}
    }
    \subfigure[$Z=6$]{
        \centering
        \includegraphics[width=0.38\textwidth]{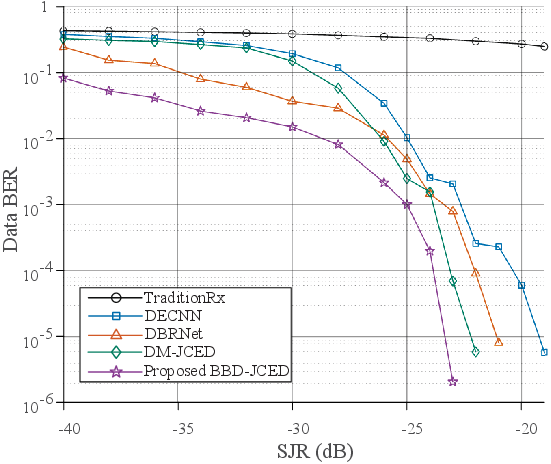}
    }
    \caption{Data BER performance under LFM jamming with different numbers of sweep periods $Z$.}
\label{fig:eval_lfm}
\end{figure*}

Fig. \ref{fig:eval_lfm} shows the corresponding results under LFM jamming with different numbers of sweep periods $Z$, while retaining the TDL-A channel model and an SNR of $20$ dB. TraditionRx again exhibits a substantial BER floor, whereas BBD-JCED achieves the lowest BER for both $Z=4.5$ and $Z=6$, with an approximately $1\sim5$ dB SJR advantage over \emph{DECNN}, \emph{DBRNet}, and \emph{DM-JCED} around a target BER of $10^{-5}$. Increasing $Z$ from $4.5$ to $6$ generally makes reliable recovery more challenging and shifts the BER transition toward higher SJR values. Despite this change in the time-frequency jamming pattern, BBD-JCED preserves the lowest BER among the tested schemes. These results indicate that explicitly constructing a Brownian bridge between the jamming-suppressed OFDM observation and the target data representation, together with refined CSI and jamming-mask guidance, provides more reliable recovery under the same jamming-suppression front-end and sampling budget.

Table \ref{table:complexity_bbjr} compares the inference complexity of all receiver schemes. All schemes include the proposed jamming suppression module as a common front-end, and the reported FLOPs and parameter counts therefore cover the complete receiver pipeline. The back-end schemes use the same OFDM-grid input dimension of $(B,C,L,N_s)=(1,2,256,14)$. TraditionRx has the lowest computational cost due to its conventional processing structure. Among the learning-based receivers, BBD-JCED requires $355.9\times10^{6}$ FLOPs and $7.8\times10^{4}$ parameters, reducing FLOPs by approximately $14.0\%$, $44.9\%$, and $18.2\%$ relative to \emph{DECNN}, \emph{DBRNet}, and \emph{DM-JCED}, respectively. Together with the BER results, BBD-JCED achieves improved bit recovery performance with competitive inference complexity.

\begin{table}[!ht]
\caption{Inference complexity of tested receiver schemes}
\label{table:complexity_bbjr}
\centering
\scalebox{0.9}{
\begin{tabular}{c|c|c|c}
\toprule
\textbf{Method} &\textbf{\makecell{Input Dimension \\ $(B,C,L,N_s)$}} &\makecell{\textbf{FLOPs} \\($\times 10^{6}$)} &\makecell{\textbf{Parameters}\\ ($\times 10^4$)}\\
\midrule
TraditionRx &\multirow{5}*{(1,2,256,14)} &\textbf{33.6} &\textbf{1.4} \\
DECNN &  &413.6 &35.6 \\
DBRNet &  &645.7 &12.9  \\
DM-JCED & &435.1 &18.9  \\
Proposed BBD-JCED & &\textbf{355.9} &\textbf{7.8}  \\
\bottomrule
\end{tabular}
}
\end{table}

\section{Conclusion}
\label{conclusion}

In this paper, we proposed a Brownian bridge diffusion-assisted joint channel estimation and data detection framework for jamming-resilient OFDM receivers, with particular emphasis on the joint degradation of pilot and data symbols caused by time-frequency jamming. The proposed framework integrates a jamming suppression module with a Brownian bridge-joint recovery module. The former identifies jamming-contaminated time-frequency bins in the STFT domain and suppresses the dominant jamming components through spectral notching. Based on the resulting suppressed OFDM observation, the latter performs the reverse recovery process to reconstruct the coded bits. To characterize this recovery process probabilistically, we established a Brownian bridge-based probabilistic graph in which the refined CSI and estimated spectral notching mask are incorporated as physical side information, and derived a VLB to reformulate the intractable bit estimate problem into a tractable training objective. Furthermore, a two-stage training algorithm was developed, in which the jamming suppression module is first optimized and subsequently frozen, while the Brownian bridge-joint recovery module is trained using the VLB-guided objective. Simulation results under representative CSN and LFM jamming scenarios demonstrated that the proposed BBD-JCED framework achieves improved jamming suppression and BER performance over the considered baseline schemes while maintaining competitive inference complexity.

\appendices
\section{Proof of Theorem \ref{theorm:vlb}}\label{appendix:vlb}

When regarding $p(\mathbf{X}_{t_{0:T-1}}|\mathbf{b},\mathbf{X}_{t_T})$ as the variational distribution and applying Jensen's inequality, the VLB of $\ln p(\mathbf{b}|\mathbf{X}_{t_T},\mathbf{\Theta})$ can be expressed as
\begin{equation}
    \begin{aligned}
    &\ln p(\mathbf{b}|\mathbf{X}_{t_T},\mathbf{\Theta})\\
    \ge&\underbrace{\mathbb{E}_{p(\mathbf{X}_{t_{0:T-1}}|\mathbf{b},\mathbf{X}_{t_T})}\left[\ln\frac{p(\mathbf{b},\mathbf{X}_{t_{0:T-1}}|\mathbf{X}_{t_T},\mathbf{\Theta})}{p(\mathbf{X}_{t_{0:T-1}}|\mathbf{b},\mathbf{X}_{t_T})}\right]}_{=:\text{VLB}}.
    \label{eq:vlb_init}
\end{aligned}
\end{equation}

\begin{figure*}[b]
\hrule
\begin{equation}
    \begin{aligned}
    p(\mathbf{X}_{t_{1:T-1}}|\mathbf{X}_{t_0},\mathbf{X}_{t_T})&=p(\mathbf{X}_{t_{T-1}}|\mathbf{X}_{t_0},\mathbf{X}_{t_T})\prod_{\eta=2}^{T-1} p(\mathbf{X}_{t_{\eta-1}}|\mathbf{X}_{t_\eta},\mathbf{X}_{t_0},\mathbf{X}_{t_T}),\\
    p_{\phi}(\mathbf{X}_{t_{0:T-1}}|\mathbf{X}_{t_T},\mathbf{\Theta})&=p_{\phi}(\mathbf{X}_{t_{T-1}}|\mathbf{X}_{t_T},\mathbf{\Theta})\prod_{\eta=1}^{T-1} p_{\phi}(\mathbf{X}_{t_{\eta-1}}|\mathbf{X}_{t_{\eta}},\mathbf{X}_{t_T},\mathbf{\Theta}).
    \end{aligned}
    \label{eq:chain}
\end{equation}

\hrule

\begin{equation}
    \begin{aligned}
    \text{VLB}&=-\sum_{\eta=2}^{T}\mathbb{E}_{p(\cdot|\mathbf{b},\mathbf{X}_{t_T})}\left[D_{\text{KL}}(p(\mathbf{X}_{t_{\eta-1}}|\mathbf{X}_{t_\eta},\mathbf{X}_{t_0},\mathbf{X}_{t_T})||p_{\phi}(\mathbf{X}_{t_{\eta-1}}|\mathbf{X}_{t_\eta},\mathbf{X}_{t_T},\mathbf{\Theta}))\right]\\
    &\quad\,+\mathbb{E}_{p(\cdot|\mathbf{b},\mathbf{X}_{t_T})}\left[\ln p_{\psi}(\mathbf{b}|\mathbf{X}_{t_0},\mathbf{\Theta})\right]+\mathbb{E}_{p(\cdot|\mathbf{b},\mathbf{X}_{t_T})}\left[\ln p_{\phi}(\mathbf{X}_{t_0}|\mathbf{X}_{t_1},\mathbf{X}_{t_T},\mathbf{\Theta})\right]-\mathbb{E}_{p(\cdot|\mathbf{b},\mathbf{X}_{t_T})}\left[\ln p(\mathbf{X}_{t_0}|\mathbf{b})\right].
    \label{eq:vlb_kl}
\end{aligned}
\end{equation}
\end{figure*}

Based on \eqref{eq:transition_markov}, we can obtain the decomposition in \eqref{eq:chain}. Combining it with \eqref{eq:decomp}, the VLB term in \eqref{eq:vlb_init} can be further expressed as \eqref{eq:vlb_kl}. 

Combining \eqref{eq:transition_forward_state}, \eqref{eq:transition_markov} and Bayes' theorem, the conditional distribution $p(\mathbf{X}_{t_{\eta-1}}|\mathbf{X}_{t_\eta},\mathbf{X}_{t_0},\mathbf{X}_{t_T})$ in \eqref{eq:vlb_kl} can be expressed as
\begin{equation}
    \begin{aligned}
    &p(\mathbf{X}_{t_{\eta-1}}|\mathbf{X}_{t_\eta},\mathbf{X}_{t_0},\mathbf{X}_{t_T})\\
    =&\frac{p(\mathbf{X}_{t_{\eta}}|\mathbf{X}_{t_{\eta-1}},\mathbf{X}_{t_T})p(\mathbf{X}_{t_{\eta-1}}|\mathbf{X}_{t_0},\mathbf{X}_{t_T})}{p(\mathbf{X}_{t_{\eta}}|\mathbf{X}_{t_0},\mathbf{X}_{t_T})}\\
    =&\mathcal{N}\left(\mathbf{X}_{t_{\eta-1}};\tilde{\boldsymbol{\mu}}(\mathbf{X}_{t_\eta},\mathbf{X}_{t_0},\mathbf{X}_{t_T}),\tilde{\delta}_{t_{\eta-1}|t_{\eta}}\mathbf{I}\right),
    \end{aligned}
\label{eq:transition_reverse_truth}
\end{equation}
with $\tilde{\boldsymbol{\mu}}(\mathbf{X}_{t_\eta},\mathbf{X}_{t_0},\mathbf{X}_{t_T})$ and $\tilde{\delta}_{t_{\eta-1}|t_{\eta}}$ expressed as
\begin{equation}
    \begin{aligned}
    \tilde{\boldsymbol{\mu}}(\mathbf{X}_{t_\eta},\mathbf{X}_{t_0},\mathbf{X}_{t_T})&=\frac{t_{\eta-1}}{t_{\eta}}\mathbf{X}_{t_{\eta}}+\left(1-\frac{t_{\eta-1}}{t_{\eta}}\right)\mathbf{X}_{t_0},\\
    \tilde{\delta}_{t_{\eta-1}|t_{\eta}}&=\frac{t_{\eta-1}(t_{\eta}-t_{\eta-1})}{t_{\eta}}.
    \end{aligned}
\label{eq:transition_reverse_truth_mean}
\end{equation}

Substituting \eqref{eq:transition_reverse}, \eqref{eq:transition_reverse_mean}, \eqref{eq:transition_reverse_truth}, and \eqref{eq:transition_reverse_truth_mean} into \eqref{eq:vlb_kl}, the VLB can be simplified as
\begin{equation}
    \begin{aligned}
    &\text{VLB} \\
    \propto& -\sum_{\eta=1}^{T}C_{\eta}\mathbb{E}_{(\mathbf{X}_{t_0},\mathbf{X}_{t_T},\mathbf{\Theta})}\left[\left\|\mathcal{V}_{\phi}(\mathbf{X}_{t_\eta},\mathbf{X}_{t_T},\mathbf{\Theta},t_\eta)-\mathbf{X}_{t_0}\right\|_F^2\right]\\
    &+\mathbb{E}_{(\mathbf{b},\mathbf{X}_{t_0})}\left[\ln p_{\psi}(\mathbf{b}|\mathbf{X}_{t_0})\right],
    \label{eq:vlb_simp}
\end{aligned}
\end{equation}
where $\{C_\eta\in\mathbb{R}_{>0}\}$ are constants independent of the learnable parameters $\phi$ and $\psi$.


 




\vfill

\end{document}